\documentclass[aps,twocolumn,showpacs,preprintnumbers,amsmath,amssymb,prl]{revtex4-1}
\usepackage{epsf,amsmath,amssymb,verbatim,color,multirow,pifont}
\usepackage{graphicx}

\newcommand{\nubar}{{\bar \nu}}

\begin{document}

\title{Finite-size scaling theory for explosive percolation transitions}
\author{Y.S.~Cho$^1$, S.-W.~Kim$^2$, J.D.~Noh$^{2,3}$, B.~Kahng$^{1,3}$
and D.~Kim$^{1,3}$}
\affiliation{{$^1$ Department of Physics and Astronomy, Seoul
National University, Seoul 151-747, Korea}\\
{$^2$ Department of Physics, University of Seoul, Seoul 130-743,
Korea}\\
{$^3$ School of Physics, Korea Institute for
Advanced Study, Seoul 130-722, Korea}}
\date{\today}

\begin{abstract}
The finite-size scaling (FSS) theory for continuous phase transitions has been
useful in determining the critical behavior from the size dependent behaviors of
thermodynamic quantities. When the phase transition is discontinuous, however,
FSS approach has not been well established yet. Here, we develop a FSS theory
for the explosive percolation transition arising in the Erd\H{o}s and
R\'enyi model under the Achlioptas process. A scaling function is derived
based on the observed fact that the derivative of the curve of the order
parameter at the critical point $t_c$ diverges with system size in a power-law
manner, which is different from the conventional one based on the divergence
of the correlation length at $t_c$. We show that the susceptibility is also described
in the same scaling form. Numerical simulation data for different system sizes are well
collapsed on the respective scaling functions.
\end{abstract}

 \maketitle

Explosive percolation transition (PT) occurring in a modified Erd\H{o}s and
R\'enyi (ER) model has attracted considerable attention in physics communities
in a short period~\cite{science,ziff,cho,santo,herrmann1,souza,herrmann2,friedman},
because most studies of discontinuous phase transitions have been
limited to equilibrium thermal systems. The modified ER model
includes an additional rule that discourages the formation of a giant cluster,
called the Achlioptas process (AP)~\cite{science} to the rule in the
classic ER network model~\cite{er}.
According to this rule, while a giant cluster develops slowly, many large-size
clusters accumulate with their numbers exceeding to those in the critical state.
After an enforced delay, a giant cluster is formed at a transition point by the
aggregation of the large-size clusters, which proceeds in an extremely short time.
Thus, the order parameter increases suddenly. In analogy to the eruption
of a volcano or a seismic outbreak, this transition is called the explosive PT.

Finite-size scaling (FSS) theory has been useful for characterizing phase transitions.
When the phase transition is continuous, the critical behavior of a system in the
thermodynamic limit can be extracted from the size dependent behaviors of
thermodynamic quantities. For example, the magnetization of ferromagnetic systems
is assumed to follow a scaling form near the critical temperature $T_c$,
\begin{equation}
m=L^{-\beta/\nu}f((T-T_c) L^{1/\nu}),\label{finite_scaling}
\end{equation}
where $L$ is the lattice size, $f(x)$ is a scaling function,
and $T$ is temperature. $\beta$ and $\nu$ are the critical exponents
associated with the magnetization and the correlation length, respectively.
This scaling function was set up based on the fact that the correlation
length, which diverges as $T\to T_c$, is limited to the lattice size $L$.
When the phase transition is discontinuous, however, the correlation length
does not diverge. Thus, a new approach was needed.
As a successful example, a scaling function was derived by using the probability
distribution for the internal energy for the $q$-state Potts model
with $q > 4$ in two dimensions~\cite{challa,binder}.
However, FSS approach for discontinuous transitions arising in disordered systems
has not been studied yet. In this Letter, we develop FSS theory
for the discontinuous PT in the modified ER model under the AP,
and obtain a scaling function, which has a different form from
the conventional one (\ref{finite_scaling}).

In the classic ER model, a system is composed of a fixed number of vertices $N$,
which evolves as one edge is randomly added to it at each time step.
Hereafter, time is defined as the number of edges added to each node.
In the ER model under the AP, at each time step, two edges are randomly selected,
but only one of them is actually added to the system to minimize the
product of the sizes of the clusters that are connected by the potential edge.
The ER model based on this product rule is hereafter called the ERPR model.
While the rule in the ERPR model
may be too artificial to fit real systems, the ERPR model contains an intrinsic
mechanism triggering discontinuous phase transitions. Thus, it is meaningful
to develop a FSS theory with this ERPR model.

\begin{figure}[t]
\includegraphics[width=0.8\linewidth]{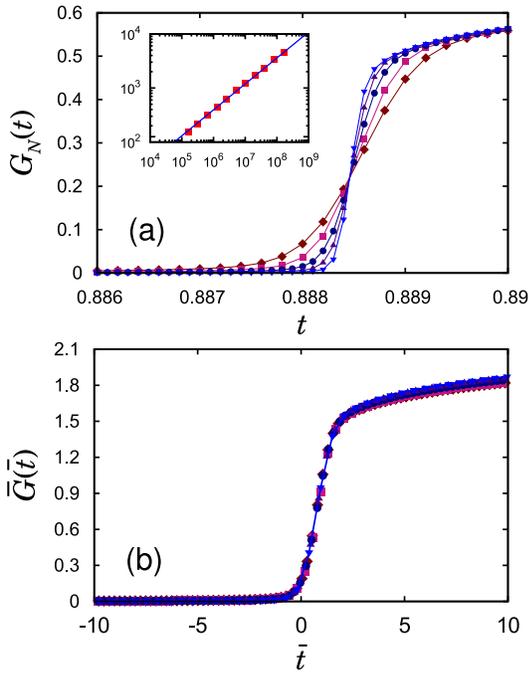}
\caption{(Color online) (a) Plot of the giant cluster size per node $G_N(t)$
versus time $t$ for system sizes $N/10^4=128,$ 256, 512, 1024 and 2048.
The curves intersect at $t_{\rm x}$, which is regarded as the transition
point in the thermodynamic limit. Inset: Plot of $dG_N(t)/dt$ at $t_{\rm x}$
versus $N$. The solid line is a reference line with a slope of $0.48$.
(b) The same plot with scaled variables
${\bar G}({\bar t})\equiv r^{-1}G({\bar t})$ and ${\bar t}\equiv (t-t_d)r^{-1}
\frac{dG_N}{dt}|_{t_{\rm x}}$, where $r=M_2^2(t_d)/M_3(t_d)$.
Data for different system sizes are well collapsed in a single
curve.}\label{giant}
\end{figure}

The size of the giant cluster is determined as $G_N(t)\equiv 1-\sum_s^{\prime} sn_s(t)$,
where $n_s(t)$ is the density of $s$-size clusters at time $t$, and the largest cluster is excluded in the sum, denoted by the prime in the summation.
Fig.\ref{giant} shows $G_N(t)$ versus $t$ for different system sizes for the ERPR model. The curves of $G_N(t)$ for different $N$s intersect at approximately one point,
namely, $t_{\rm x}\approx 0.88845(2)$. We consider the $t$-intercept of the tangent of the curve $G_N(t)$ at $t_{\rm x}$, denoted as $t_d(N)$. Then this time is calculated as
\begin{equation}
t_d(N)=t_{\rm x}-G_N(t_{\rm x})\Big(\frac{dG_N(t)}{dt}\Big|_{t=t_{\rm x}}\Big)^{-1}.\label{time_relation}
\end{equation}
We find that as $N\to \infty$, the derivative of $G_N(t)$ at $t_{\rm x}$ diverges as
\begin{equation}
\frac{dG_N(t)}{dt}\Big|_{t=t_{\rm x}}\sim N^{\theta},\label{slope}
\end{equation}
with $\theta \approx 0.48(1)$.
Thus, the derivative diverges in the limit $N\to \infty$, indicating that
the PT is indeed discontinuous.

\begin{figure}[t]
\includegraphics[width=0.8\linewidth]{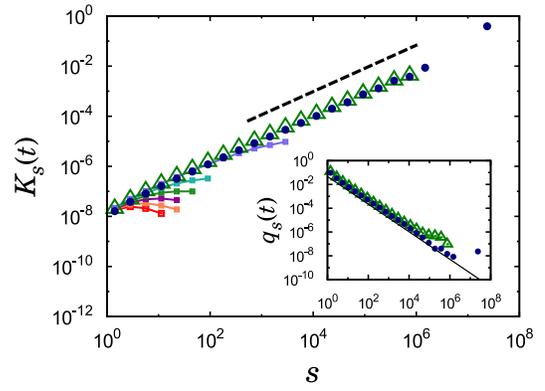}
\caption{(Color online) Plot of $K_s(t)$ versus $s$ at various
times for the ERPR network with system size $N=8.192 \times 10^7$. The dashed line is a
reference line with a slope of 0.92. The data correspond to $t=0.4$, 0.5, 0.6,
0.7, 0.8, 0.88, $t_d(\triangle)$, and $t_z(\circ)$ (starting from the bottom).
Inset: Plot of $q_s(t_d)(\triangle)$ and $q_s(t_z)(\circ)$ versus $s$. The solid line
is a reference line with a slope of $-1.16$.}\label{kernel}
\end{figure}

Since the dynamics of the ERPR model involves the selection of two
edges at each time step, it may be difficult to analytically clarify the evolution
of the giant cluster. Thus, a simple model, called the cluster-aggregation
network model, which involves the selection of one edge at each time step, was
introduced~\cite{can}. This cluster-aggregation network model is a simple
representation of the ERPR model, because the frequency of intercluster
connections is dominant until the percolation threshold.
In this model, once two clusters of sizes $i$ and $j$
are selected with probabilities $i^{\omega}/c(t)$ and $j^{\omega}/c(t)$,
respectively, where $c(t)\equiv \sum_s s^{\omega} n_s(t)$,
one vertex is chosen randomly from each selected cluster,
and the two vertices are connected. When $\omega=1$,
this model reduces to the ER network.
To put the ERPR model in this perspective, we measure corresponding
value of $\omega$ by measuring the probability $q_s(t)$
of an added edge being connected to a vertex in a cluster of size $s$ at time
$t$. This probability is given by $q_s(t)=s^{\omega}n_s(t)/c(t)$ for the
cluster-aggregation network model. Thus, by measuring $q_s(t)$ and $n_s(t)$,
we calculate $s^{\omega}/c(t)$ for the ERPR model, which is denoted as $K_s(t)$.
When $t \ll t_d(N)$, $K_s(t)$ does not exhibit a power-law behavior,
however, as $t\to t_d(N)$, it does as $K_s(t) \sim s^{\omega}$ with $\omega\approx 0.92$ (Fig.~\ref{kernel}). This power-law behavior persists as time
progresses beyond $t_d(N)$.

Since $\omega_{\rm ERPR}\approx 0.92$ is close to $\omega_{\rm ER}=1$,
we may get a hint on the evolution of the ERPR model in the context of the
analytic formulae of the ER model. The explicit formula for
the evolution of the giant component of the ER model with an arbitrary initial
condition was calculated~\cite{ziff2,can}. For example, the giant cluster
size near the transition point $t_c=\frac{1}{2M_2(0)}$ is
\begin{equation}
G^{\rm ER}(t)=\frac{2M_2^2(0)}{M_3(0)}\left(\frac{t}{t_c}-1\right),\label{giant_er}
\end{equation}
where $M_n(0)=\sum_s^{\prime} s^n n_s(0)$ is the initial $n$-th moment.
Thus, when a system starts an evolution from $n_s(0)=\delta_{s,1}$, thus
$M_2(0)=M_3(0)=1$, the giant cluster size begins to grow from zero
to a macroscopic scale continuously at the transition point $t_c=1/2$.

Here, we consider a particular initial condition, in which $M_2(0)$
and $M_3(0)$ depend on $N$. We assume that $n_s(0)$ follows
a flat distribution, $n_s(0)=n_0$, in the range $0 < s < s_{m}$,
where $s_m$, the size of the largest cluster at $t=0$, depends on
$N$ as $s_{m}=N^{\eta}$. Then, $n_0=2N^{-2\eta}$, $M_2(0)\propto
N^{\eta}$, and $M_3(0)\propto N^{2\eta}$. Then, a PT takes place
at $t_c(N)=1/2M_2(0)\propto N^{-\eta}$, and $G(t)\sim
r(2M_2(0)t-1)$ for $t > t_c(N)$ from Eq.~(\ref{giant_er}), where
$r \equiv 2M_2^2(0)/M_3(0)$ turns out to be in $\mathcal{O}(1)$.
Thus, if time $t$ is scaled as $t^{\prime}=tM_2(0)$,
then $G(t^{\prime})$ has a mean field behavior similar to the
original ER case. This scaling behavior implies that
$t$ increases $\sim \mathcal{O}(N^{-\eta})$ beyond $t_c$,
$G_N(t)$ does so by $\mathcal{O}(1)$.
Thus, a discontinuous phase transition occurs.
Generally, {\it if the conditions that (i) the amplitude $M_2^2(0)/M_3(0)$
is finite and (ii) $M_2(0)$ diverges as $N\to \infty$ are fulfilled,
a discontinuous PT can occur in the ER model.} We show below that the
origin of the explosive PT in the ERPR model can be understood
in this scheme.

For further discussion, we reconsider the ER model with an initial condition
$n_s(0)=\delta_{s,1}$. As time passes, small-size clusters develop,
and thus the cluster-size distribution is no longer monodisperse. We take
a certain time $t^* < t_c$ and the amplitude at that time is
$r(t^*)=2M_2^2(t^*)/M_3(t^*)$. Since the evolution of the ER model
proceeds continuously, it holds that $G^{\rm ER}(t)=r(t^*)[2M_2(t^*)(t-t^*)-1]$,
in which the certain time $t^*$ is regarded as an {\it ad hoc} time origin.
This relation has been checked numerically. We apply this logic to
the evolution of the ERPR model below.

\begin{figure}[t]
\includegraphics[width=0.8\linewidth]{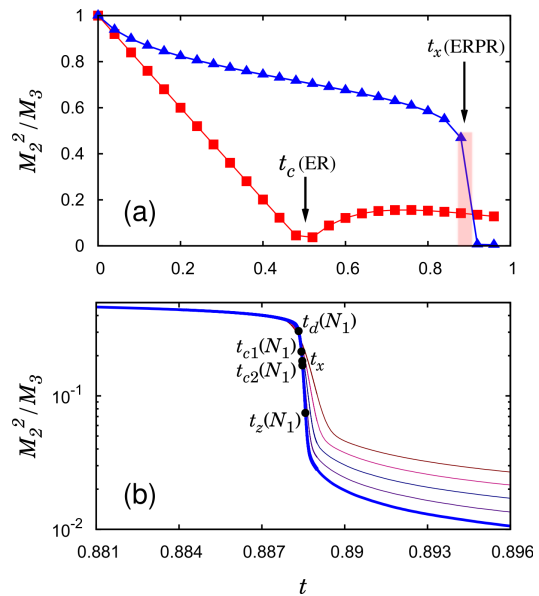}
\caption{(Color online) (a) Plot of the amplitude $M_2^2(t)/M_3(t)$
versus time $t$ for the ERPR network ($\triangle$, blue) with system size
$N_1=2.048 \times 10^{7}$ and for the ER model ($\square$, red).
(b) Zoom-in plot of the amplitude
versus $t$ for the ERPR network with different system sizes $N=1.28 \times
10^6$, $2.56\times 10^6$, $5.12 \times 10^6$ $1.024 \times 10^7$ and
$2.048\times 10^7$ from the top. The amplitude decreases more drastically
as system size is increased.
Various characteristic times are indicated on the curve for $N=N_1$.
The explosion starts at $t_d(N)$ and ends at $t_z(N)$.
The two kinds of susceptibility reach their maximum at
$t_{c1}$ and $t_{c2}$, respectively,
while the crossing time $t_x$ is independent of $N$.}
\label{amplitude}
\end{figure}

We examine the amplitude $M_2^2(t)/M_3(t)$ as a function of time
for the ERPR model near the transition point, and find that it is
finite but suddenly drops to zero in the thermodynamic limit
(Fig.~\ref{amplitude}(a)). In the conventional ER model, however,
$M_2^2(t)/M_3(t)$ is zero near the transition point, and thus,
the transition is continuous (Fig.~\ref{amplitude}(a)).
Interestingly, the time $t_d(N)$ defined in Eq.(\ref{slope})
locates approximately on the verge of the sudden drop in the amplitude
(Fig.~\ref{amplitude}(b)). Thereby, we call $t_d(N)$ as the triggering
time of the explosive PT. We examine the behavior
of the amplitude and the second moment at $t_d(N)$ as a function of system
size $N$ more carefully. We find the amplitude can be fit to a power law
$M_2^2(t_d)/M_3(t_d)\sim N^{-\kappa}$. However, we find that $\kappa$ is
very small and decrease slowly from $0.06$ to $0.04$, measured by
successive slopes, as the system size increases from $10^5$ to $2\times
10^7$. Thus, one may regard $\kappa$ as zero. This suggests that the amplitude
is independent of $N$ in the thermodynamic limit.
However, $M_2(t_d(N))\sim N^{0.7}$. Next, we take $t_d(N)$ as the {\it ad hoc}
time origin, and then $t_d(N)$ plays a role of the percolation threshold
as we have seen that $t_c(N)\to 0$ as $N \to \infty$ in the ER model
with the flat distribution. Indeed,
$t_d(N)$ approaches $t_{\rm x}$ as in (\ref{time_relation}). From these
results, we can conclude that the ERPR model satisfies the condition for a
discontinuous phase transition at the {\it ad hoc} time origin.
On the other hand, it was noticed~\cite{friedman} that the key ingredient
to explosive transition was not the details of the dynamics during
the actual explosion, but rather lay in the period preceding the
explosion when a type of power keg developed. This ingredient is
embodied in the moments of the cluster size distribution at $t_d(N)$,
which satisfies the conditions (i) and (ii).
Thus, the explosive PT takes place.

We derive a scaling function for the ERPR model, inspired by the formula
(\ref{giant_er}). Since $\omega^{\rm ERPR}\ne \omega^{\rm ER}$, we take
a general form of the giant cluster size, $G(t)=r(ut-1)$,
where $r$ and $u$ can depend on $N$ and $t$.
Since we take $t_d(N)$ as the {\it ad hoc} time origin, we calculate
$r$ and $u$ at $t_d(N)$, and substitute $t \to t-t_d(N)$.
Then, we obtain that $u(t_d)=r^{-1}(t_d)\frac{dG_N}{dt}|_{t_{\rm x}}$,
which is not $M_2(t_d)$ as in the ER case. We assume that $r(t_d)$ follows
the same form $M_2^2(t_d)/M_3(t_d)$ from the ER model.
Thus, our FSS form becomes $G_N(t)=r(t_d)\bar{G}(\bar{t})$,
where $\bar{G}(x)$ is a scaling function and $\bar{t}\equiv (t-t_d)r^{-1}\frac{dG_N}{dt}|_{t_x}$.
Then,
\begin{equation}
G_N(t)= N^{-\beta/\nubar} f_0\big((t-t_d(N))N^{1/\nubar}\big),
\label{scaling_form}
\end{equation}
where $f_0(x)$ is a scaling function and $\beta/\nubar=\kappa$ and
$1/\nubar=\theta+\kappa$.
We confirm this scaling function with numerical data obtained
for different system sizes as shown in Fig.~\ref{giant}(b).
Even though we obtain $\kappa\approx 0.06 - 0.04$ in our
simulation range of system size, we expect $\beta/\nubar=0$ asymptotically
and thus $1/\nubar \approx 0.48(1)$ and $\beta=0$. We notice that the scaling
function is in a particular form: Instead of following the conventional form
$t-t_{\rm x}$ in its argument, the scaling function in (\ref{scaling_form})
contains the form $t-t_d(N)$, implying that the interval $t-t_d(N)$
can change with $N$ for a fixed $t$.

\begin{figure}[t]
\includegraphics[width=1.0\linewidth]{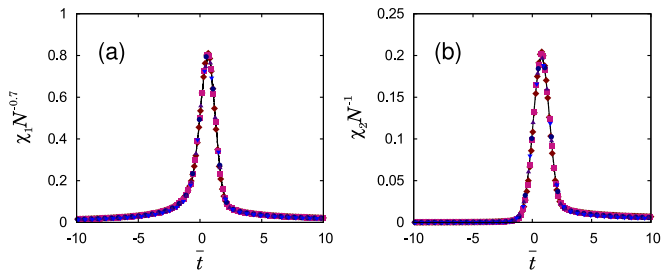}
\caption{(Color online) (a) Plot of the susceptibility $\chi_1$, the mean
cluster size, versus
time in a scaled form for different system sizes used in Fig.~\ref{giant}.
${\bar t}\equiv (t-t_d)r^{-1}
\frac{dG_N}{dt}|_{t_{\rm x}}$, where $r \equiv M_2^2(t_d)/M_3(t_d)$.
(b) Similar plot for the susceptibility $\chi_2$, the fluctuation of
the giant cluster size, versus time in a scaled form.}
\label{susc}
\end{figure}

We define the susceptibility in two ways. The first is the mean cluster size
$\chi_1 (t)\equiv \sum_s^{\prime} s^2 n_s(t) / \sum_s^{\prime} s n_s(t)$,
which exhibits a peak at $t_{c1}(N)$. $t_{c1}(N)$ is larger than
both $t_d(N)$ and $t_{\rm x}$, but approaches $t_x^+$ as $N\to \infty$.
The susceptibility at $t_{c1}(N)$ increases with $N$ as
$\chi_1(t_{c1}(N))\sim N^{0.7}$, confirming the previous result~\cite{santo2},
which also behaves in the same manner at $t_d(N)$.
Thus, $\chi_1(t)$ is written in the form $\chi_1(t)=N^{\gamma_1/\nubar}f_1\big((t-t_{d})N^{1/\nubar}\big)$,
where $f_1(x)$ is another scaling function. Thus, $\gamma_1/\nubar \approx 0.7$.
The second one is the fluctuation of the giant component sizes defined as
$\chi_2(t) \equiv N\sqrt{\langle G_N^2(t) \rangle-\langle G_N(t) \rangle^2}$.
This quantity exhibits a peak at $t_{c2}$. We find that
$\chi_2(t_{c2}(N))\sim N$, which holds even at $t_d(N)$.
Thus, $\gamma_2/\nubar=1$ and $\chi_2(t)=N f_2\big((t-t_d)N^{1/\nubar}\big)$
with a scaling function $f_2$. The scaling behaviors are confirmed numerically
in Fig.~\ref{susc}.

\begin{figure}[t]
\includegraphics[width=0.8\linewidth]{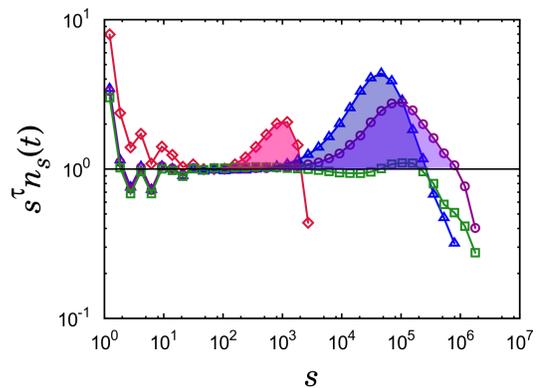}
\caption{(Color online) Plot of $s^{\tau}n_s(t)$ versus $s$ for different
times $t=0.884(\diamond)$, $t_d(\triangle)$,
$t_{c1}(\circ)$ and $t_z(\square)$ for the ERPR model
with size $N=1.024 \times 10^7$.
The hump size, the number of nodes contained in the shade area, becomes maximum
at time $t_d(N)$, beyond which it reduces.}\label{hump}
\end{figure}

To catch the physical meaning of $t_d(N)$ further, we investigate the cluster
size distribution $n_s(t)$ as a function of time $t$.
In early times, $n_s(t)$ decays exponentially. As time passes, it exhibits a
power-law behavior in small cluster-size region, but a hump develops in the
tail region. We represent $n_s(t)$ as $n_s(t)\sim s^{-\tau(t)}g(s,t,N)$,
where the exponent $\tau(t)$ is weakly dependent on time $t$ and ranges from
$\tau\approx 2.05(2)- 2.09(1)$ as time progresses beyond $t_d(N)$.
$g(s,t,N)$ represents the shape of the hump. To estimate the hump size, we first
plot $n_s(t)s^{\tau}$ versus $s$ for different times and define the hump size
as the number of nodes in the shade region in Fig.\ref{hump}.
As time passes, the shape area increases and attains a maximum at $t_m(N)$,
and then it reduces to zero at time $t_z(N)$. In fact, $t_z(N)$ is defined
as the point at which the size distribution of finite clusters follows
a power law. We obtain the value of $\tau$ at $t_z$ as
$\tau\approx 2.09(1)$~\cite{self}. The characteristic time $t_m(N)$ is
numerically consistent with $t_d(N)$, indicating that indeed $t_d(N)$
is the triggering time of the explosive PT, and the powder keg becomes empty
at $t_z(N)$.

In summary, we have found that the curves of the giant component
size $G_N(t)$ for different system sizes $N$s intersect at the transition
point $t_{\rm x}$, and the derivative
of $G_N(t)$ with respect to $t$ at $t_{\rm x}$ diverges in a
power-law manner as $N$ increases.
This indicates that the explosive PT is manifestly discontinuous.
We also found that the evolution can be regarded as kinetic cluster aggregations
in which the connection kernel turns out to be similar to that for the conventional
ER model near the transition point. Based on these facts, we have developed a
finite-size scaling theory for for the behavior of the explosive PT, which has a different
origin from the conventional one for continuous phase transitions, and thus
has a different form. We have determined the critical exponents.
The method we develop here can also be used for other percolating systems
exhibiting discontinuous transitions, for example, in the interacting complex
networks~\cite{havlin}.

This study was supported by an NRF grant awarded through the Acceleration Research Program (Grant No. 2010-0015066) (BK), the NAP of KRCF (DK) and the Seoul Science Foundation (YSC).

\vfil\eject

\begin{thebibliography}{99}
\bibitem{self} The measured value $\tau\approx 2.09$ is related to $\omega\approx 0.92$ by the relationship $\tau=3-\omega$ but with logarithmic correction, which can be shown analytically.
\end{thebibliography}
\end{document}